# Cryptography and Key Management Schemes for Wireless Sensor Networks

Jaydip Sen
Department of Data Science
Praxis Business School, Kolkata, India
email: jaydip.sen@acm.org.

**Abstract**

Wireless sensor networks (WSNs) are made up of a large number of tiny sensors, which can sense, analyze, and communicate information about the outside world. These networks play a significant role in a broad range of fields, from crucial military surveillance applications to monitoring building security. Key management in WSNs is a critical task. While the security and integrity of messages communicated through these networks and the authenticity of the nodes are dependent on the robustness of the key management schemes, designing an efficient key generation, distribution, and revocation scheme is quite challenging. While resource-constrained sensor nodes should not be exposed to computationally demanding asymmetric key algorithms, the use of symmetric key-based systems leaves the entire network vulnerable to several attacks. This chapter provides a comprehensive survey of several well-known cryptographic mechanisms and key management schemes for WSNs.

## 1. Introduction

Wireless sensor networks (WSNs) are made up of a large number of tiny sensors, which can sense, analyze, and communicate information about the outside world. These networks play a significant role in a broad range of fields, from crucial military surveillance applications to monitoring building security [1]. In these networks, a sizable number of sensor nodes are placed throughout a big field, where the operational environment is frequently hostile or severe, to monitor it. However, because of their low processing speed, little memory, and insufficient energy, WSN nodes face significant resource limitations. Hence, these networks need to include security features to protect against attacks like physical tampering, node capture, denial of service, eavesdropping, etc. as they are typically placed in distant locations and left unattended.

Unfortunately, resource-constrained sensor nodes cannot implement typical security measures because of their large overhead. Researchers in WSN security have put out many security protocols that are tailored to these networks' resource limitations. Researchers in WSN security have proposed several protocols for secure and efficient routing [2-5], securely aggregating data for protecting data privacy [6-11], etc.

Since WSN architectures are mostly decentralized, and due to the lack of infrastructure, security procedures used in WSNs need also to incorporate cooperation among the nodes and address more security challenges like secure routing and aggregation of data. In the real-world deployment scenario, WSNs cannot be a priori taken to be reliable. To address the issues that standard cryptographic algorithms are unable to address, researchers have concentrated on developing a sensor trust model [12-24].

Vulnerability to physical attacks is a significant concern in WSNs since the sensor nodes are typically unattended and physically unsafe. There are several ideas in the literature for protecting sensor nodes from physical attack [25-34].

The choice of the cryptographic scheme and the key distribution and management protocol for a WSN is an extremely critical decision as the entire security of the network is based on these schemes. However, designing a computationally efficient yet highly secure key management scheme is a challenging task. While these resource-constrained sensor nodes should not be exposed to computationally demanding public key-based algorithms, the use of symmetric key cryptography leaves the network vulnerable to several attacks. This chapter provides a comprehensive survey and a comparative analysis of various cryptographic mechanisms and key management schemes in the current literature.

The rest of the chapter is organized as follows. Section 2 presents different cryptographic schemes used in WSNs including the public key and the symmetric key-based algorithms and systems. Section 3 discusses several key management schemes including the network architecture-based protocols and deterministic, and probabilistic key distribution mechanisms. Finally, Section 4 concludes the chapter and highlights some future research directions.

## 2. Cryptographic Schemes for WSNs

In WSNs, choosing the best cryptographic technique is essential since cryptography provides all security functions. The code size, data size, processing time, and power consumption of cryptographic techniques used in WSNs should all be taken into consideration together with the sensor node limits. We concentrate on the choice of cryptography in WSNs in this section. We first discuss public key cryptography, then delve into systems that use symmetric keys for their cryptographic functions.

### 2.1 Public Key Cryptographic Mechanisms in WSNs

Many experts think that public key protocols such as the Diffie-Hellman key exchange [35] or RSA [36], should not be used in WSNs because of the code complexity, data size, processing time, and power consumption these algorithms involve.

A single security operation typically requires dozens or even millions of multiplication instructions, which makes public key methods like RSA computationally demanding. Furthermore, the number of CPU cycles needed to execute an instruction for the multiplication operation is a critical factor in determining a microprocessor's efficiency for a public key method [37].

In resource-constrained wireless devices, Brown et al. discovered that public key methods like RSA typically take some minutes to execute cryptographic operations such as encryption and decryption. This is a long enough time for an adversary to launch denial of service (DoS) attacks [38]. Carman et al. observed that a basic 128-bit operation of multiplication often requires thousands of nano-joules from a microprocessor [37].

As opposed to public key methods, the algorithms of hash functions and symmetric keys involve substantially lower processing overhead. An AES block of 128-bit size typically consumes an energy of 0.104 mJ, which is substantially lower than the anticipated energy consumption for a 1024-bit block when utilizing RSA on the MC68328 DragonBall CPU [37].

By employing the appropriate choice of parameters in the algorithms and optimized approaches that consume lower power for execution, research has demonstrated that it is possible to deploy public key-based protocols in WSNs [39-41]. Elliptic Curve Cryptography (ECC) [42,43], Ntru-Encrypt [44], RSA [36], and Rabin's Scheme [45] are some of the public key algorithms that have been studied for this purpose. The RSA and ECC algorithms are the subjects of most studies in the literature. ECC is appealing because it is highly secure even with smaller keys. Hence, the use of ECC decreases the requirement of processing and transmission costs. While RSA with 1024-bit keys offers a degree of security that is currently acceptable for many applications, the same level of security is achieved using ECC with a 160-bit key (ECC-160) [46]. As per the new recommendation, a key size of 2048 bits is used in the RSA protocol as the minimum size of the key. This is like the 224-bit ECC protocol [47].

On an Atmel ATmega128 CPU, Wander et al. evaluated the amount of energy required in RSA and ECC protocols for authentication and key exchange [41]. The Elliptic Curve Digital Signature Algorithm (ECDSA) generates and verifies the ECC-based signature [48]. The handshake in the secured socket layer (SSL) requires two entities: a client that initiates the session, and a server that responds to the request [49]. The key exchange scheme is a more compact form of this handshake. Each sensor in the WSN is presumed to have a certificate that has been signed using the private key of the trusted authority. The two parties validate their respective certificates during the handshake phase and agree on the session key that will be used for communication. The findings indicate that compared to RSA signatures, ECDSA signatures are much less expensive. Additionally, on the server side, the ECC protocol has superior performance, while the RSA protocol performs better on the client side. However, the two protocols do not exhibit any significant difference in the power requirement in carrying out the key exchange operation. Additionally, as the key size grows, ECC outperforms RSA in terms of relative performance.

The use of encryption operations in RSA and ECC on Mica2 motes demonstrated the viability of the use of public key protocols in WSNs [50]. The design of the TinyPK system proposed by Watro et al. uses the TinyOS development environment to build the RSA system on Mica2 motes [51]. The authors have shown that this technique effectively implements authentication and key agreement protocol in sensor nodes with limited resources. Another ECC-based technique called TinyECC [52] has been created and put into use on Mica2. Malan et al. also carried out similar work using ECC on Mica2 [40]. A single symmetric key was distributed via ECC for the TinySec module's link-layer encryption.

While sensor nodes could be able to perform public key cryptography, the cost of private key operations remains high. In some cases, the [40,50] assumptions might not be true. For instance, [40,50] solely focused on the public key activities, presuming a base station or outside party would handle the private key operations. The operation time of the public key may be made to be very quick by choosing the right parameters, for instance, by utilizing the tiny number $e = 2^{16} + 1$ as the public key, while the operation time of the private key remains constant. Several public key operations are not available in this framework due to the restriction of operations using private keys exclusively at a base station. Peer-to-peer authentication and secure data are two examples of such services.

## 2.2 Symmetric Key Cryptography in WSNs

As symmetric key cryptography approaches involve less computational overhead than public key cryptographic mechanisms, most research studies for WSNs concentrate on their utilization. A single shared key between the two communicating hosts is employed by symmetric key cryptographic techniques and is used for both encryption and decryption. But efficiently and securely distributing a common key to two nodes for secure communication is a significant barrier to the widespread use of symmetric key encryption. Given that it might not always be possible to pre-distribute the key, this is a challenging topic.

Five well-known encryption techniques were tested on six different microprocessors, with word sizes ranging from 8 bits (Atmel AVR), 16 bits (Mitsubishi M16C), and 32 bits (StrongARM, XScale) in [53]. These included RC4 [54], RC5 [55], IDEA [54], SHA-256 [56], and MD5 [54,57]. For each algorithm and platform, execution time and code memory size were assessed. The studies showed that each encryption class and architectural class had a consistent cryptographic cost. While support for the Instruction Set Architecture (ISA) is only confined to certain impacts on specific protocols, the influence of caches was minimal. Additionally, hashing techniques (like MD5 and SHA-1) are found to consume more resources in comparison to RC4 and IDEA encryption algorithms.

Law et al. studied the performance of the RC5 and TEA symmetric key algorithms in [58]. On the MSP430F149 from IAR Systems, six additional block ciphers are also assessed [58]. These block ciphers are Rijndael, Camellia, KASUMI, RC6, and RC5. The benchmarking criteria were CPU cycles, data RAM, and code.

For WSN security services to be provided, the proper cryptography mechanism for sensor nodes must be chosen. The capability of the sensor nodes for calculation and transmission, however, determines the outcome. Hardware design and encryption algorithms are both active areas of study.

As mentioned earlier, studies have observed the viability of public key-based protocols in WSNs even if they have higher resource requirements. Private key operations can still not be completed in a sensor node due to the high computational and energy costs involved. Further research is needed on the use of operations using symmetric keys shared among the nodes in a WSN. In terms of speed and low energy consumption, symmetric key cryptography is preferable to public key cryptography. However, key distribution methods using shared symmetric keys are not flawless. Designing effective and adaptable key distribution strategies is necessary. To meet the growing demands on computing and communication in sensor nodes, it is also anticipated that stronger motes will need to be developed.

## 3. WSN Key Management Protocols

Key management has gotten the most attention from researchers studying WSN security. A crucial strategy for ensuring network services and application security in WSNs is key generation, storage, and distribution. Establishing keys for the nodes efficiently and securely is the main objective of a key management scheme. The key management system should allow for network node insertion and revocation. These methods must be extremely lightweight even for a large-scale network, due to the power and memory constraints of the nodes. Due to their high computational overhead, the public key cryptographic approaches do not find many applications in WSNs. Most of the protocols for key management are based on the use of shared keys. Figure 1 depicts a classification of the currently existing key management schemes for WSNs. The works discussed in this section are from impactful publications in the WSN literature from 2000 to 2022.

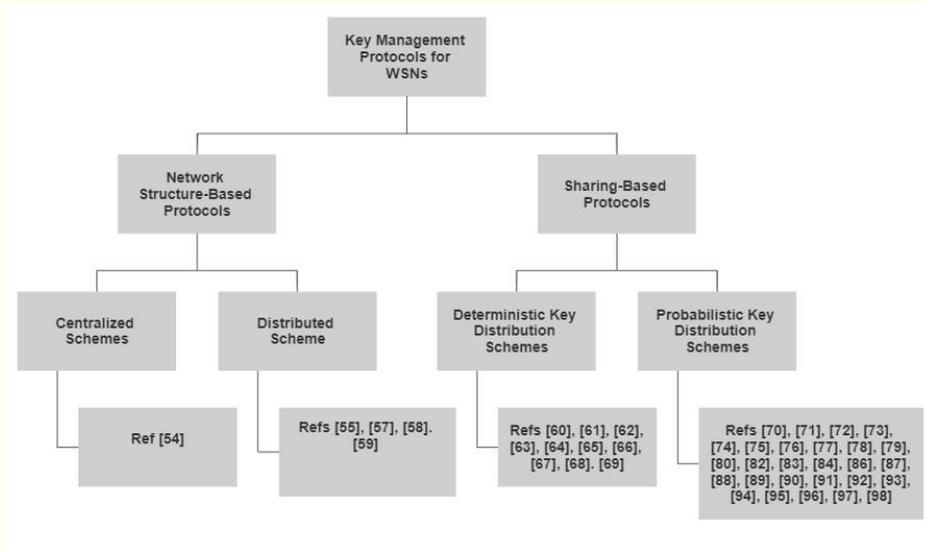

**Figure 1:** The classification of key management schemes for WSNs

### 3.1 Network Architecture-Based Key Management Schemes

The distribution or centralization of the key management task depends on the underlying architecture of the network. The production, distribution, and revocation of keys are all under the control of one entity under a centralized key management scheme. The name of this organization is Key Distribution Centre (KDC). The LKHW method is a protocol for WSNs that uses a single key distribution in a centralized manner [59]. The basis of the LKHW protocol is the hierarchy of logical keys. The hierarchy of the keys leads to a tree structure in which the base station is at the root position of the tree. The base station plays the role of the KDC in the network. This scheme's sole point of failure is its biggest flaw. The whole network's security will be compromised if the central controller malfunctions. Another problem is that it cannot be scaled. Additionally, it doesn't offer data authentication. Different controllers are used in the distributed key management protocols to manage key-related tasks. These protocols enable higher scalability and do not have a single point of failure problem. The majority of key management techniques that have been studied so far are dispersed in nature.

To secure the sensor network, Qin et al. [60] proposed an approach that involves building an AVL tree [61] for key management along with the use of elliptic curve cryptography (ECC) [41]. The AVL tree stores each node's public key and the identifier of its neighboring nodes. The scheme is efficient in several aspects, including processing overhead, memory space requirement, and overhead of communication. *elliptic curve pallier encryption* (ECPE), a cryptographic technology, is also used in this strategy to defend against numerous security risks. Another element of this strategy was constantly updated keys.

A scheme proposed by Swaminathan et al. [62] uses the topology of a wireless network and creates a structure aggregating several distributed spanning trees (DSTs). The proposed scheme, known as the *efficient low-cost key generation mechanism* (ELWKM), is found to be involving low overhead in computation and memory requirement.

An efficient public key cryptography-based strategy was presented by Chen et al. [63]. The scheme combined the Merkle hash tree, the Bloom filter, and several other encryption and decryption techniques. The elliptic curve discrete logarithm issue makes use of key threshold theory to create a key management system.

For clustered WSNs, Yao et al. presented a key management method known as the *local key hierarchy* (LKH++) [64]. A dynamically constructed tree is used for storing the keys in the nodes of the network. For secure communication among a group of nodes in a cluster, the keys are used for encryption and authentication. The sink node i.e., the base station stores and manages the tree. When needed for the network, this method regenerates and rekeys the keys. The LKH++ scheme provides a WSN with increased robustness against several attacks.

These methods, which can be classified as deterministic or probabilistic, are covered further in this section.

## 3.2 Sharing-Based Key Management Schemes

The likelihood of the availability of a shared key between any two randomly chosen nodes in a WSN is used as a basis for the classification of the key management techniques. The essential management strategies might be either deterministic or probabilistic depending on this likelihood.

### 3.2.1 Deterministic Key Distribution Schemes

Zhu et al. proposed a protocol for key distribution protocol in WSNs [65,66]. The scheme, known as the *localized encryption and authentication protocol* (LEAP), is based on cryptographic operations using shared keys among the nodes. Depending on the security needs of each packet, it employs a separate keying scheme. Each node is assigned one of four different types of keys: (i) a unique pre-distributed key shared between the nodes and the sink node, (ii) a set of keys shared among the nodes of the network, (iii) keys shared among neighboring node pairs, and (iv) a shared key among all members of a cluster. Peer-to-peer communication is secured using the pair-wise keys shared with nearby nodes, and local broadcast is secured using the cluster key.

The time needed to launch an attack on a node is longer than that needed for the network to build. A node will be able to discover all its intermediate neighbors during this period. Each node is deployed with a shared initial key already loaded. A master key is generated for each node based on their shared key and the individual identity of the node. Then, sensor nodes communicate by exchanging hello messages. The hello messages are verified by the recipients (the neighbor's master key may be calculated because the shared key and identification are known). Based on their master keys, the nodes then compute a shared key. After the setup, the common key is deleted in every node, and it is assumed that no node has been hacked thus far. Injecting bogus data or decoding messages sent earlier is now very difficult as no attacker can get access to the shared key. No node may afterward fabricate the master key of another node, either. This establishes the shared keys for all node pairs amongst all neighboring nodes located nearby. A node creates the cluster key after the keys for the node pairs are generated. With the help of the shared key between a node pair, the cluster is derived and the cluster key is delivered in an encrypted form to all the neighboring nodes. The group key is installed in the nodes a priori, and it is revoked and regenerated as soon as a compromised node is found. In a crude method, the sink node may communicate the new shared key to each cluster member node using its unique key, or it can do it one hop at a time using cluster keys. For the same, more complex algorithms have been developed. The authors have also offered strategies for creating shared keys amongst multi-hop neighbors.

A broadcast session key (BROSK) negotiation protocol has been put out by Lai et al. [67]. The master key used by BROSK is assumed to be shared by all network nodes. Node A in the WSN sends a broadcast message to its neighbor node B for initiating the creation of a shared session key. A shared session key is eventually agreed upon by the two nodes. The protocol is found to exhibit high scalability and power efficiency.

Utilizing combinatorial design theory, Camete & Yener presented a key generation mechanism for sensor nodes in a connected network [68]. Block design approaches in combinatorics are the foundation of the key generation strategy that utilizes *combinatorial design theory* (CDTKeying). Methods such as generalized quadrangle and symmetric design are used for this purpose.

The method creates a symmetric design with the following parameters: $n^2 + n + 1$, $n + 1$, 1. It does this by using a projective plane of a finite order $n$, i.e., for prime powers of $n$. The system employs a key pool that has the size of $n^2 + n + 1$ and supports $n^2 + n + 1$ nodes. It creates $n^2 + n + 1$ key chains of size $n + 1$, each key appearing in precisely $n + 1$ number of key chains, and each pair of chains of keys sharing exactly one key. Each pair of nodes discovers precisely one key common to them after deployment. Hence, there is no chance of the existence of a shared key between any node pair. The need requirement of n being prime is a shortcoming of this claim. As a result, a given key chain size can accommodate all network sizes.

Two deterministic methods based on combinatorial design theory were suggested by Lee and Stinson: the *deterministic multiple spaces Bloms' scheme* (DMBS) and the *ID-based one-way function scheme* (IOS) [69]. In [70], they went into further detail on how combinatorial set systems may be used to create deterministic key pre-distribution methods for WSNs.

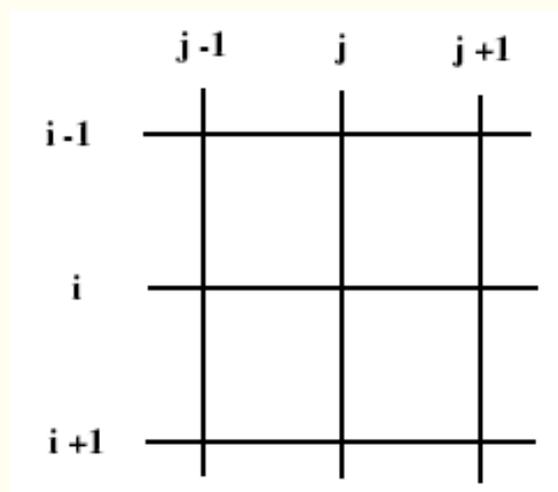

**Figure 2:** The grid structure for node placement in the PIKE protocol [66]

A deterministic key management approach has been proposed by Chan and Perrig [71]. The proposed scheme is based on the generation of pair-wise shared keys among the neighboring nodes in a WSN. A novel technique, *peer intermediaries for key establishment* (PIKE), is utilized for arranging all the $N$ sensor nodes in a network in the form of a 2-D space as shown in Figure 2, with each node's coordinate being $(x, y)$ where, $x, y \in \{0, 1, ..., \sqrt{N} – 1\}$. There are $2(\sqrt{N} - 1)$ nodes with identical $x$ or $y$ coordinate values while each of these nodes has distinct pair-wise keys. An intermediary node that has one of its coordinates identical to both nodes is designated dynamically as the intermediate router. The role of the router is to route the key from two nodes that do not share a common coordinate. However, the safe connectivity of the scheme is only 2 / √N. This implies that every node should generate a key for all its neighboring nodes possibly utilizing multi-link routes. As a result, the communication overhead of the method will be significantly high.

A *hybrid authenticated key establishment* (HAKE) technique that makes use of the computational and energy differences between a sensor node and the base station in a WSN has been put forth by Huang et al. [72]. The authors contend that a single sensor node has far less computational and energy capacity than a base station. Hence, the main cryptographic computations are delegated to the central node (i.e., the base station). Lightweight symmetric-key procedures are used on the sensor side. Elliptic curve cryptography is used by the base station and sensors to authenticate each other. In the suggested technique, a public key's validity is additionally verified using certificates. The elliptic curve scheme is the foundation for the certificates. These certificates can be used to confirm the legitimacy of sensor nodes.

A $t$-degree $(k + 1)$-variate symmetric polynomial is used in Zhou and Fang's *scalable key agreement scheme* (SKAS), which is a deterministic key agreement methodology for generating keys in a WSN [73].

Gandino et al. [74] proposed a key management scheme for WSNs that involves the generation of a master key. The scheme is known as the *random seed distribution with transitory master key* (RSDTMK). The master key is further used in combination with a puzzle in generating the shared keys among the nodes. The shared keys are used in establishing secure communication between any pair of nodes in the network.

### 3.2.2 Probabilistic Key Distribution Schemes

The majority of key distribution techniques used in WSNs are based on distributed, probabilistic systems. A *random key pre-distribution* (RKPD) approach for WSNs has been presented by Eschenauer and Gligor [75]. It is based on probabilistic key sharing between random network nodes. Key pre-distribution, shared key discovery, and path key establishment are the three stages of the mechanism. Each sensor has a key ring installed in it during the key pre-distribution step. A wide collection of $P$ keys is randomly selected to create the $k$ keys that make up the key ring. The base station also keeps track of the associations between the sensor identification and the key IDs on the key ring. A pair-wise key is shared by each sensor node and the base station. Every node identifies its neighboring node with whom it has a shared key during the phase of shared key discovery. For this, the authors proposed two strategies. The basic approach involves every node broadcasting a list of plaintext key IDs from their key rings and enabling nearby nodes to determine whether those nodes have any shared key with the node. However, an attacker can use this method to track the pattern of key sharing among the nodes. The advanced approach, unlike the basic strategy, conceals key-sharing patterns between nodes from an attacker by using the challenge-response methodology. After the second phase, a path key is finally allocated in the path key setup phase for those nodes that are within the range of communication but do not have any shared key among them. The base station can instruct all nodes to revoke the keys in the king ring of a node if the node is found to be compromised. The key revocation process is identical to the key regeneration process. For authenticating the messages from the base station, the shared keys between the base station and the nodes are used. This defends against any possible attempt of a base station impersonation attack. If a node is hacked, the likelihood of an attacker successfully attacking any connection is around $k/P$. It only has an impact on a few sensor nodes because, $k << P$. This key distribution mechanism is considered to be the fundamental method among the random key distribution techniques in WSNs. There have been several other major pre-distribution strategies put forth [76-81].

Any two neighbor nodes in the basic random key management scheme must locate a single shared key from their key rings to create a safe link during the key configuration phase. However, Chan et al. found that raising the key ring's degree of key overlap might improve the network's resistance to node capture [77]. The authors' suggested a pre-distribution approach *for q-composite random keys* (QCRK). To create a safe link between any two neighbor nodes, they must share at least *q* common keys during the key establishment step. To improve the fundamental random key management technique, they also included a key update step. Let's say that following the key establishment step, *A* and *B* have a secure link, and the secure key is *k* from the key pool *P*. The security of the link between *A* and *B* is at risk if any of those nodes are taken over since *k* could be stored in the keyring memory of some other nodes in the network. The communication key between *A* and *B* should thus be updated rather than utilizing a key from the key pool. The authors have included a *multi-path key reinforcement* for the key update as a solution to this issue. If an opponent wishes to recover the communication key in this scenario, he or she must listen in on every disjoint link that connects nodes *A* and *B*. An additional layer of security is added to the system by using a *random pair-wise key management* approach for node-to-node authentication.

Typically, both nodes must broadcast their key indices or use a challenge-response mechanism to uncover common keys to determine whether the key sets of two nodes cross. Such techniques involve a significant communication overhead. By connecting a node's key indices and identification, Di Pietro et al. [79] proposed an *extended random key distribution* (ERKD) system. For instance, the key indices for each node are calculated as $g(ID, i)$ for $i$ = 1, 2, ..., N, where *ID* is the node identity. Each node is given a *pseudo-random number generator*, denoted by $g(x, y)$. By confirming its node identification, other nodes can determine which key is in its key set.

Du et al. proposed a *deployment knowledge-based random key distribution* (DKRKD) system that makes use of deployment knowledge of WSNs and avoids irrelevant key assignments [80]. The authors contended that in many practical cases, some deployment knowledge may be accessible a priori which may be gainfully exploited in designing a key distribution protocol. The proposed protocol is found to significantly enhance the performance of WSNs and made the networks robust against adversarial attacks.

A polynomial-based key predistribution strategy for *group key pre-distribution* (GKPD) that may be used for WSNs was presented by Blundo et al. [81]. The bivariate *t*-degree polynomial presented in (1) is generated at random by the key setup server.

$$f(x,y) = \sum_{i=0}^{t}\sum_{j=0}^{t} a_{ij}\, x^i y^j$$

(1)

The bivariate polynomial is generated $\mathbf{F}_q$ over a finite field *q*, where *q* is a prime big enough to hold a key for cryptography. By selecting $a_{ij} = a_{ji}$ a symmetric polynomial, $f(x, y) = f(y, x)$, is obtained. It is expected that every sensor node has a distinct, integer-valued, non-zero identity. Each sensor node *u*, which has a share of polynomial $f(u, y)$ is loaded with the coefficients of the polynomial $f(u, y)$. Nodes *u* and *v* broadcast their IDs when they need to create a shared key. By computing $f(u, y)$ at $y = v$, node *u* may then derive $f(u, v)$, and node *v* can compute $f(v, u)$ by evaluating $f(v, y)$ at $y = u$. The shared key between nodes *u* and *v* has been determined to be $K_{uv} = f(u, v) = f(v, u)$ because of the polynomial symmetry. A bivariate polynomial of degree *t* is also $(t + 1)$-secure. To reconstruct the polynomial, an adversary must compromise at least $(t + 1)$ nodes that have the same key shares.

A polynomial *pool-based key pre-distribution* (PPKP) strategy has been put out by Liu et al. [78]. Additionally, there are three stages to the scheme: setup, direct key establishment, and path key creation. The setup server creates a set $F$ of bivariate $t$-degree polynomials over the finite field $\mathbf{F}_q$ at random during the setup phase. The setup server selects a subset of polynomials $F_i \subseteq F$ for each sensor node and allocates the polynomial shares of these polynomials to node $i$. The sensor nodes locate a common polynomial with other sensor nodes during the direct key formation step and then create a pair-wise key using the polynomial-based key predistribution strategy described in [81]. The phase of the path key establishment is comparable to that of the fundamental random key management method. The key predistribution schemes based on random subset assignment and the grid-based key predistribution scheme are also described and analyzed in the paper. Additionally, the suggested framework enables the investigation of many instantiations.

Blom's key predistribution approach [82] is used in Du et al.'s *multiple-space key pre-distribution* (MSKP) system [76]. The system in [78] is based on a set of bivariate $t$-degree polynomials, whereas the scheme in [76] is based on Blom's approach. This is the main distinction between the schemes presented in [76] and [78]. The suggested approach enables any pair of network nodes to locate a pair-wise secret key. The network is completely safe as long as no more than λ nodes are attacked. The base station then generates a random (λ +1) x $N$ matrix $G$ over a finite field $GF(q)$, and an $N$ x (λ + 1) matrix $A = (D.G)^T$, where $(D.G)^T$ is the transpose of the matrix $D.G$. Matrix $D$ must be maintained a secret and must not be revealed to attackers. It is simple to confirm that $A.G$ is a symmetric matrix using (2).

$$A.G = (D.G)^T.G = G^T.D^T.G = G^T.D.G = (A.G)^T$$

(2)

Hence, $K_{ij} = K_{ji}$. $K_{ij}$ (or $K_{ji}$) is intended to serve as the pair-wise key connecting nodes $i$ and $j$. The two procedures listed below are completed in the pre-distribution phase for any sensor node $k$ to perform the aforementioned computation: The $k$-th column of matrix $G$ and the $k$-th row of matrix $A$ are both stored at node $k$, respectively. The pair-wise key between nodes $i$ and $j$ must then be determined. To do this, nodes $i$ and $j$ swap their private rows of $A$ before computing $K_{ij}$ and $K_{ji}$, respectively. Each sensor node in the proposed approach is loaded with $G$ and $\tau$ unique $D$ matrices that are selected from a sizable pool of $\omega$ symmetric matrices $D_1$, $D_2$, ....., $D_\omega$ of dimension (λ + 1) x (λ + 1). The $j$-th row of $A_i$ should be stored at this node after computing the matrix $A_i = (D_i.G)^T$, for each $D_i$. Each node must determine whether it shares any space with neighbors after deployment. If the nodes discovered that they shared a space, they could use Blom's approach to create a pair-wise key. The plan is adaptable and scalable. Additionally, compared to the plan put forth in [78], it is far more durable to node capture.

A *lightweight polynomial-based key management* (LPKM) strategy for distributed WSN was put out by Fan et al. [83]. In addition to providing secure one-to-one and many-to-one communications using polynomial-based keys (such as the pairwise key, cluster key, and group key), this protocol also provided authentication using a probabilistic local broadcast authentication protocol among nearby nodes.

To provide the security of personal key shares, Wang et al. [84] presented a *hash-chain-based key management* (HCKM) strategy that was inspired by polynomials. It employs $p$-degree polynomial $F(x)$ to provide safe communication between and within classes. Consider a sensor network with two groups, $G_1$ and $G_2$, with the first group being $G_1$. If a member of group $G_1$ uses the key $P(v)$ to encrypt the multicast message for group $G_2$ members. The group controller gives a polynomial to each member of groups $G_1$ and $G_2$ so they may use it to decode this message using the key $P(x)$ that members of group $G_2$ obtained from members of group $G_1$. A revocation polynomial and a specific one-way hash function are utilized in this key distribution scheme's defense against the *collusion* attack. The one-way hash chain technique of generating the revocation polynomial is used to update the broadcast transmission. This strategy reduces communication costs and eliminates the collusion attack.

A key management system based on *polynomials by self-healing keys* (PSHK) has been presented by Sun et al. [85]. The enhanced polynomials and broadcast authentication technique can offer collision resistance and secure communication. The pairwise keys between the controller node and other sensor nodes are produced using a collection of sliding windows and enhanced polynomials. *Sch-I* and *Sch-II*, two distinct strategies, were also put forth. The *Sch-I* technique puts forward the notion that the controller node and other sensors establish and share pairwise keys. *Sch-I* may be dynamically updated in response to the network. *Sch-I* rejects the vulnerability since other nodes are unaware of this polynomial. A one-way hash function provides *forward security*, whereas a *modified polynomial* provides *backward security*. *Sch-II* enhances security by removing the hash chain. By using this approach, they were able to increase *collision resistance* while avoiding the drawbacks of acceding polynomials.

Chebyshev polynomials [86] are a novel key management strategy that Ramkumar et al. [87] have used to create keys for the nodes. To protect message communications, the proposed scheme, known as *key management using Chebyshev polynomial* (KMCP), utilizes the features of Chebyshev's polynomials.

A novel, efficient, and *dynamic key management* (DKM) strategy for sensor networks was presented by Zhou et al. [88]. To create effective keys, a mix of trivariate symmetric polynomials, ECC, and *p*-degree polynomials were used. The key is dynamically updated using a time slice approach. The communication overhead in the key distribution scheme is minimized by using a one-way hash chain among the nodes.

Jing et al. present a *fully homomorphic encryption-based key generation* (FHEKG) scheme [89]. It produced paired keys using *homomorphic encryption* [90]. The network is protected against node capture attempts using this strategy. These pairs of keys are strong, random, and unique thanks to the characteristics of an asymmetric polynomial, which satisfies the criteria of a suitable key management method.

To create paired keys among sensor nodes, Zhan et al. suggested a system using an *equation-based key distribution* (EKD) [91]. The sensor network communicated and delivered messages discreetly using these paired keys. There is only one solution to every equation in the set of equations. The generated keys are, therefore, compact, effective, and robust. To create private shared keys, linear equations' cutting points are employed. In sensor networks, these paired keys are used to defend the network from different threats. To avoid the high computational overhead involved in solving polynomial equations, this technique generates keys and implements key management in the network using linear equations with just two variables using the *exclusion basis system* (EBS). The benefit of this strategy is that, in contrast to other conventional key schemes, it offers a solid key setup, and other performance measures are unaffected.

Dinkar et al. proposed a key distribution scheme that is based on *symmetric polynomials* using a multivariate framework [92]. In this proposition, known as the *hybrid key management security scheme* (HKMSS), the keys shared between the central node (i.e., the sink node) and the cluster heads are derived using symmetric polynomials and matrices. A secure network is created using the protocol for future communications between the nodes. The matrices are regularly updated and stored at the sink node and the cluster heads. The matrices are updated whenever the shared key between a pair is changed. The key management scheme is found to be efficient even when the shared keys are frequently updated.

To ensure that the network remains connected always, the probability of a pair of nodes having a shared key should be carefully chosen and each sensor node has to store a variety of key materials. When sensor nodes have limited memory, this results in significant storage overhead. By lowering the quantity of key-related data that must be saved in each node and assuring a specific likelihood of key sharing between a pair of nodes, an improvement over the random key distribution scheme [75] is proposed by Hwang and Kim [93]. Instead of securing connections throughout the whole network, they plan to do it in the biggest subcomponent. The likelihood that two nodes share a key is decreased, but it is still high enough to link the largest network component.

The fundamental random key management technique was expanded by Hwang et al., who also put out a *cluster key grouping* (CGA) approach [94]. The authors also proposed optimization of memory, energy requirement, and the level of security.

The essential components are evenly dispersed over the network's terrain in each of the key management systems that have been previously addressed. Because of the homogeneous distribution, the likelihood of secure connectivity—the sharing of a direct key by two neighboring nodes—is rather low. As a result, the creation of indirect keys will always include significant communication overhead. Two close sensor nodes can be preloaded with the same set of essential elements if the location of one of them is known. Secure connectivity might be enhanced in this way.

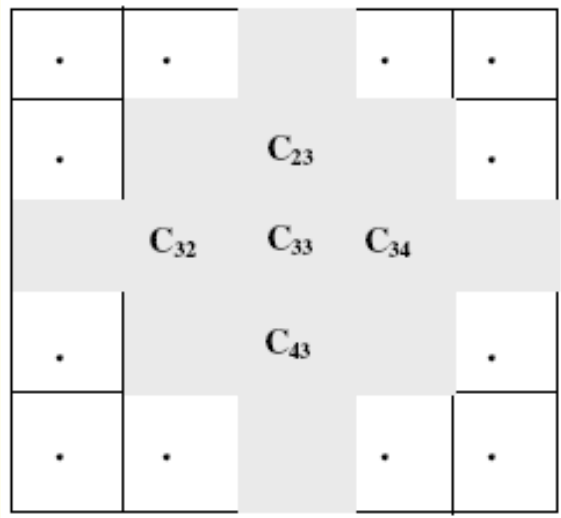

**Figure 3:** The topology of a WSN divided into several cells in the LBKP scheme [95]

Liu & Ning proposed the *location-based key pre-distribution* (LBKP) scheme, in which a WSN is split into several cells of square shapes [95]. Every cell has a certain t-degree polynomial associated with it involving two variables. The polynomials of each sensor node's home cell and four cells that are both horizontally and vertically adjacent to it are pre-loaded onto each node. Two neighbor nodes can create a shared key between them after deployment if the two nodes possess a share of the same polynomial. As an illustration, the polynomial of cell $C_{33}$ in Figure 3 is likewise allocated to cells $C_{32}$, $C_{34}$, $C_{23}$, and $C_{43}$. Other cells' polynomials are allocated similarly. A node in $C_{33}$, therefore, shares some polynomial information with other nodes in the shaded regions.

Younies et al. proposed a key distribution scheme that utilizes the location information of the nodes in a WSN [96]. This technique generates keys using a location and an *exclusion-based system* (EBS). The produced keys are pairwise, randomized, and unique and are computed based on the locations of the nodes. The proposed scheme is also referred to as the *scalable, hierarchical, efficient, location-based, and lightweight* (SHELL) protocol. This technique provides key regeneration and enhances network security against various threats including node compromise and hijacking. All of the nodes share the burden of key management, hence reducing storage overhead and compute complexity. Additionally, the scheme avoids the overload on the base station. The location information of the nodes is used to derive the shared keys between the node pairs. The scheme is resistant to node collusion attacks. SHELL offers protection from collusion attacks. These key generation and distribution strategies allow for changes in network size, such as the addition or removal of nodes, as well as key refreshes that take node location into account.

Choi et al. presented the *location-dependent key management* (LDKM) scheme for key generation and distribution based on the location of the nodes in a WSN [97]. In the proposed scheme, Grid-based coordinates are used in this method to create network keys. Nine data coordinates and eight neighbor coordinates are utilized. These coordinated paired keys are established during the network's first and second stages. The sequence number of every packet that a node sends is also used. This approach offers protection against several internal and external dangers.

Zhu & Zhan argue that while random key predistribution is the most efficient way of managing keys in a WSN, security, and robustness of the network are two important issues that must be addressed in such approaches [98]. The authors propose a *q-composite random key management* (QCRKM) approach that is based on the knowledge of network topology.

Shi et al. propose a key management scheme in WSNs that works on dynamic authentication of the member nodes [99]. The proposed mechanism, known as the *dynamic membership authentication and key management* (DMAKM) scheme, can authenticate nodes for accessing network resources while dynamically refreshing the keys used in the authentication. The scheme preserves the forward and backward secrecy of information in the nodes and is found to be resistant to node capture attacks [100].

Cheng et al. propose a *fast multivariate polynomial-based authentication* (FMPA) and key management scheme that can combine two important functions in a WSN, (i) generation of keys, and (ii) authenticating nodes in the network based on the generated keys [101]. The authentication function has a linear complexity with the number of nodes in the network, unlike other similar schemes most of which have quadratic complexity with the network size.

Kumar & Malik present a scheme for node authentication and key distribution for WSN that supports dynamic joining and leaving of nodes [102]. The scheme proposed by the authors, known as *dynamic key management for clustered networks* (DKMCN), is suited for clustered WSNs in which the keys generated by a central node are distributed securely to the cluster member nodes via the cluster head nodes. The performance analysis of the scheme exhibited its robustness against various attacks including the node capture attacks [100].

Li et al. proposed a model of key management that consists of two layers of a key pool [103]. In the proposed scheme, known as the *one-way associated key management* (OAKM) model, the authentication of the nodes is done in two phases increasing the robustness of the key distribution and management task.

Table 1 categorizes and compares the deterministic key distribution schemes for WSNs which were discussed in this chapter. The protocols are compared for the types of keys they involve, the level of scalability and security they provide, and the processing, communication, and memory they demand. A similar comparative analysis for the probabilistic key distribution schemes is presented in Table 2.

**Table 1:** Summary of the deterministic key distribution schemes for WSNs

| Prot Name | Ref | Master Key | Pairwise Key | Path Key | Cluster Key | Scalability | Robustness | Proc Load | Comm Load | Memory Load |
|---|---|---|---|---|---|---|---|---|---|---|
| LKHW | [59] | Yes | Yes | No | Yes | Medium | Low | Low | Low | Low |
| LEAP | [65, 66] | Yes | Yes | Yes | Yes | High | Low | Low | Low | Low |
| BROSK | [67] | Yes | Yes | No | No | High | Low | Low | Low | Low |
| CDTKeying | [68] | No | Yes | No | No | High | High | Medium | Medium | Medium |
| IOS & DMBS | [69,70] | No | Yes | No | No | High | High | Medium | Medium | High |
| PIKE | [71] | No | Yes | Yes | No | High | Low | Low | Low | High |
| HAKE | [72] | No | Yes | No | Yes | High | High | Medium | Medium | Medium |
| SKAS | [73] | No | Yes | No | No | High | High | Low | High | Low |
| RSDTMK | [74] | Yes | Yes | No | No | High | High | High | High | High |

**Table 2:** Summary of the random key distribution schemes for WSNs

| Prot Name | Ref | Master Key | Pairwise Key | Path Key | Cluster Key | Scalability | Robustness | Proc Load | Comm Load | Memory Load |
|---|---|---|---|---|---|---|---|---|---|---|
| Basic RKPD | [75] | No | Yes | Yes | No | High | High | Medium | Medium | High |
| MSKP | [76] | No | Yes | No | No | High | High | Medium | Medium | High |
| QCRK | [77] | No | Yes | Yes | No | High | High | Medium | Medium | High |
| PPKP | [78] | No | Yes | No | No | High | High | Medium | Medium | High |
| ERKPD | [79] | No | Yes | Yes | No | High | High | High | High | High |
| DKRKD | [80] | No | Yes | No | No | High | High | Medium | Medium | Medium |
| GKPD | [81] | No | Yes | No | Yes | Low | High | Low | High | Low |
| LPKM | [83] | No | Yes | Yes | Yes | High | High | Medium | High | Low |
| HCKM | [84] | No | Yes | Yes | Yes | High | High | Medium | High | Low |
| PSHK | [85] | No | Yes | Yes | No | High | High | High | High | High |
| KMCP | [87] | No | Yes | No | No | Medium | High | High | High | High |
| DKM | [88] | No | Yes | Yes | No | High | High | High | Low | Medium |
| FHEKG | [89] | No | Yes | No | No | Low | High | High | High | High |
| EKD | [91] | No | Yes | No | No | Low | Medium | Low | Medium | High |
| HKMSS | [92] | Yes | Yes | Yes | Yes | High | High | High | High | High |
| CKG | [94] | No | Yes | No | No | High | High | Medium | Medium | High |
| LBKP | [95] | No | Yes | No | No | High | High | Medium | Medium | Medium |
| SHELL | [96] | No | Yes | No | Yes | High | High | High | High | High |
| LDKM | [97] | No | Yes | Yes | No | High | High | High | High | High |
| QCRKM | [98] | No | Yes | Yes | No | High | High | High | High | Medium |
| DMAKM | [99] | No | Yes | No | Yes | High | High | High | High | High |
| FMPA | [101] | No | Yes | Yes | Yes | High | High | High | High | High |
| DKMCN | [102] | Yes | Yes | No | Yes | Low | High | High | High | High |
| OAKM | [103] | No | Yes | No | Yes | High | High | High | High | High |

In the following, some important challenges in designing efficient and secure key management schemes for WSNs are highlighted.

***Memory:*** A key management protocol has to satisfy two goals: high security and little overhead. Several significant establishment suggestions for sensor networks have been made, however, they seldom ever fulfill these two needs. Strong security systems often demand a lot of memory, as well as fast processors and a lot of electricity. Due to the sensor platform's hardware resource limitations, they cannot readily be supported. One bit can use more energy being transmitted than being computed in a wireless context, as is widely known. In key management protocols, indirect key establishment takes place across multi-hop communication while direct key establishment just needs one-hop communication or a few rounds of it. Highly secure communication is possible when two nodes have a high probability of establishing a direct communication link with a shared key. Multi-hop communications involve more overhead and are usually less secure. However, additional key materials are needed at each node for highly secure connectivity, which is typically impracticable, especially when the network size is huge. In light of the previous two problems, memory use might be a significant barrier when developing key management procedures for a WSN. It is crucial to figure out how to lower memory use while yet keeping a certain level of security.

***End-to-end security:*** Symmetric key cryptography's computational efficiency is one of its main advantages. Since there can be possibly many nodes in a network, it is not a good idea for each node in a network to store a shared transport layer key with each remaining node. Hence, the majority of the existing symmetric key-based systems focus on the security of the link layer. However, many WSN applications demand secure node-to-node at the transport layer. For instance, an aggregator node may combine information from several nodes and provide the aggregated result to a designated central node (or the sink node) to minimize traffic in the network. The messages communicated between the aggregator and the central node and the source nodes and the aggregator node should both be secured and privacy-protected. But in hostile circumstances, any node is vulnerable. If one of the intermediary nodes on a route is hacked, the affected node may reveal or change the message sent down the route. End-to-end security can successfully stop hostile intermediary nodes from altering messages. Public key cryptography is more costly than symmetric key technology, but it allows end-to-end security and offers flexible management. A node equipped with both public key and symmetric key-based algorithms may use the public key algorithm to generate shared keys with other nodes in WSNs. Construction and implementation of efficient and effective public key algorithms are essential for achieving this aim so that they may be extensively applied to sensor systems. Another significant issue is how to validate the validity of public keys. Otherwise, a bad node might pretend to be any other legitimate node by stealing its public key. Identity-based encryption offers a quick solution to the issue. Pairing-based ECs are frequently employed in creating symmetric keys using the identities of the nodes since the majority of identity-based cryptographic methods now in use work on elliptic curve fields. However, the pairing procedure is extremely expensive, equal to or even more so than RSA. Therefore, the primary goals for academics are to develop quick methods and implementations.

**Effective symmetric key algorithms**: Because encryption and authentication based on symmetric keys are often used in the security operations of sensor nodes, there is still a need for the development of more effective symmetric key algorithms. Each packet, for instance, must be authenticated in the link layer security protocol TinySec [104], and encryption can also be activated if important packets are transferred. As a result, symmetric key algorithms that are quick and economical should be created.

**Revocation and update of keys:** Once a shared key is established between two nodes, it may be used as a master key to create several sub-keys for various functions (authentication and encryption). Cryptanalysis over the ciphers may eventually reveal the key if it is used over a long period. It is advisable to update keys regularly to prevent cryptanalysis of the master key and those sub-keys. But picking an update interval might be challenging. It is extremely difficult to make an educated guess as to how long it will take an adversary to disclose a key through cryptanalysis since the opponents' cryptanalysis capabilities are unknown. If the key is updated after a long interval, the associated key can also be hacked by an adversary. On the other hand, if the update interval is too short, it will involve a significant overhead of computation and communication. Key revocation is an issue that is linked. A node's key must be revoked if it turns out to be malicious. Key revocation hasn't, however, been properly looked into. Even though Chan et al. [105] provided a key revocation scheme that works on key pairs generated randomly using the scheme proposed in [77]. However, the proposed distributed protocol does not generalize well and hence it is difficult to use it in combination with other key distribution schemes.

**Node compromise:** This attack may be very detrimental for WSNs. Compromised nodes can cause extremely serious harm to WSN applications and are difficult to identify since they include all the genuine key materials. It's still unclear how to prevent node compromise. Most current security protocols make an effort to limit the impact of node breaches to a narrow region by carefully designing their protocols to minimize this impact. A hardware-based strategy, though, offers more potential. With improvements in hardware design and manufacturing methods, considerably more durable, tamper-proof, and affordable devices may be mounted on WSN. These devices may cause extremely serious damage to WSN applications and cannot be readily identified. It's still unclear how to prevent node compromise. Most current security protocols make an effort to limit the impact of node breaches to a narrow region by carefully designing their protocols to minimize this impact. A hardware-based strategy, though, offers more potential. To prevent node compromise, devices that are significantly tougher, more difficult to tamper with, and less expensive can be put on the sensor platform.

## 7. Conclusion

This chapter has presented a comprehensive survey of several cryptographic and key management schemes in the current literature on WSNs. Various symmetric key and public key cryptographic approaches for WSNs have been discussed and their relative merits and demerits have been highlighted. The key management protocols were categorized into three broad categories, network topology-based schemes, deterministic key distribution schemes, and probabilistic key distribution schemes. For the network topology-based schemes, the distribution or centralization of the key management task is dependent on the network architecture. In the centralized key distribution schemes, the production, distribution, and revocation of keys are all under the control of one entity under a centralized key management scheme. However, in distributed key management schemes, the tasks of key generation, distribution, and revocation are delegated to a group of nodes in the network. While the distributed key management schemes are more secure and robust, these schemes involve higher overheads of computing, communication, and storage in comparison to their centralized counterparts. In deterministic key distribution schemes, a shared key between any randomly chosen nodes is either present or absent. On the contrary, in the probabilistic key distribution schemes, the availability of a shared key between a pair of nodes is given by probability. Several key distribution schemes of both types, deterministic, and probabilistic have been discussed in this chapter. The schemes have been compared on the types of keys used and the overhead of computation, communication, and memory, the schemes involve. Some open problems for future research directions are also discussed.